\newcommand{\be}{\begin{equation}}
\newcommand{\ee}{\end{equation}}
\newcommand{\bea}{\begin{eqnarray}}
\newcommand{\eea}{\end{eqnarray}}
\newcommand{\nn}{\nonumber\\}
\newcommand{\p}[1]{(\ref{#1})}
\newcommand{\lb}{\label}

\newcommand{\cA}{{\cal A}}
\newcommand{\cAb}{{\bar{\cal A}}}
\newcommand{\cW}{{\cal W}}
\newcommand{\cWb}{{\bar{\cal W}}}
\documentstyle[12pt]{article}
\def\theequation{\arabic{section}.\arabic{equation}}
\begin{document}
\begin{titlepage}
\begin{flushright}
LNF-01/007(P) \\
JINR-E2-2001-15 \\
hep-th/0101195\\
January 2001
\end{flushright}
\vskip 0.6truecm
\begin{center}
{\Large\bf
Towards the complete N=2 superfield Born-Infeld \break action 
with partially broken N=4 supersymmetry}
\end{center}
 \vskip 0.6truecm
\centerline{ S. Bellucci${}^{\,a,1}$, E. Ivanov${}^{\,b,2}$,
S. Krivonos${}^{\,b,3}$ }

\vskip 0.6truecm
\centerline{${}^a${\it INFN-Laboratori Nazionali di Frascati,}}
\centerline{\it P.O.Box 13, I-00044 Frascati, Italy}

\vspace{0.5cm}
\centerline{${}^b${\it Bogoliubov Laboratory of
Theoretical Physics, JINR,}}
\centerline{\it 141 980 Dubna, Moscow region,
Russian Federation}

\vspace{0.2cm}
\vskip 0.6truecm  \nopagebreak

\begin{abstract}
\noindent We propose a systematic way of constructing $N=2,\, d=4$ 
superfield Born-Infeld
action with a second nonlinearly realized $N=2$ supersymmetry. 
The latter,
together with the manifest $N=2$ supersymmetry, 
form a central-charge extended
$N=4,\, d=4$ supersymmetry. We embed the
Goldstone-Maxwell $N=2$ multiplet into an infinite-dimensional off-shell
supermultiplet of this $N=4$ supersymmetry and impose 
an infinite set of covariant constraints
which eliminate all extra $N=2$ superfields through 
the Goldstone-Maxwell one.
The Born-Infeld superfield Lagrangian density is
one of these composite superfields. The constraints can be solved
by iterations to any order in the fields. We present
the sought $N=2$ Born-Infeld action up to the 10th order. It 
encompasses the action found earlier by Kuzenko and
Theisen to the 8th order from a self-duality requirement. This is 
a strong indication that the
complete $N=2$ Born-Infeld action with partially broken $N=4$ supersymmetry
is also self-dual.
\end{abstract}

\vfill

\noindent{\it E-Mail:}\\
{\it 1) bellucci@lnf.infn.it}\\
{\it 2) eivanov@thsun1.jinr.ru}\\
{\it 3) krivonos@thsun1.jinr.ru}
\newpage

\end{titlepage}

\renewcommand{\thefootnote}{\arabic{footnote}}
\setcounter{footnote}0
\setcounter{equation}0
\section{Introduction}
For many reasons it is important to know off-shell superfield
actions of supersymmetric extensions of the Born-Infeld (BI) theory
\cite{CF}-\cite{Ke} and to
understand the geometry behind them. One of the basic sources of
interest in such actions is that their notable subclass, the
BI actions with a hidden extra nonlinearly realized supersymmetry,
provides a manifestly worldvolume supersymmetric description of various
Dp-branes in a static gauge \cite{Ts}. As was demonstrated
in \cite{BG} (see also \cite{RT}), this sort of BI actions 
supplies a nice example of
systems with partial spontaneous breaking of global supersymmetry 
(PBGS). The
covariant superfield gauge strengths in terms of which such actions are
formulated, can be identified with the Goldstone superfields supporting a
nonlinear realization of some underlying extended supersymmetry. 
The manifest
supersymmetry of the given BI action is the linearly realized half of
the underlying supersymmetry.

At present, the Goldstone superfield BI actions are known in a closed
explicit form only for the 1/2 PBGS options $N=2 \rightarrow N=1$
in $d=4$ \cite{BG,RT} and $d=3$ \cite{IK}. They
amount to the worldvolume actions of the spacetime-filling
D3- and D2-branes in a fixed gauge and involve, respectively,
the $N=1,\, d=4$ and $N=1,\, d=3$ vector multiplets as the Goldstone ones.

In \cite{{BIK1},{BIK2}} it was suggested that, 
by analogy with the construction of
ref. \cite{BG}, $N=2,\, d=4$ vector multiplet could serve as the
Goldstone multiplet for the 1/2 spontaneous breaking of $N=4,\, d=4$
supersymmetry. The associated Goldstone superfield action should be
a particular representative of the $N=2$ supersymmetric BI actions, such that
it possesses a hidden $N=2$ supersymmetry in parallel with the manifest one.
By inspection of the component field content of the $N=2$ vector multiplet,
it is obvious that such action should describe a D3-brane in $D=6$, with
the scalar component fields parameterizing two transverse directions.
The $N=2$ BI action constructed in \cite{Ke} reveals
no hidden extra supersymmetry \cite{KT} and so it can be
regarded merely as a part of the hypothetical genuine 
$N=4 \rightarrow N=2$ BI action.

In recent papers \cite{BIK0,BIK3} we showed how the full set of
superfield equations describing the $N=2 \rightarrow N=1$ BI system in $d=3$
and the $N=2 \rightarrow N=1$, $N=4 \rightarrow N=2$ and $N=8 \rightarrow
N=4$ ones in $d=4$ can be deduced from the customary nonlinear realizations
approach applied to the relevant PBGS patterns. A characteristic
common feature of these superfield systems is that the pure BI part of the
corresponding bosonic equations always appears in a disguised form 
in which the
Bianchi identity for the Maxwell field strength and the dynamical equation
are mixed in a tricky way. On the other hand, the equations 
for the scalar fields
(in the $N=4 \rightarrow N=2$ and $N=8 \rightarrow N=4$ cases in which
the Goldstone vector multiplets include such fields) come out in
a form explicitly derivable from the standard static-gauge Nambu-Goto
actions. The disguised form of the BI equations can be split into
the kinematical and dynamical parts by a nonlinear equivalence
redefinition of the corresponding bosonic component field.
As was demonstrated in \cite{BIK3} for the $N=4 \rightarrow N=2$ example,
the superfield version of this redefinition is an equivalence transformation
{}from the original basic $N=2$ Goldstone superfield to the standard 
$N=2$ Maxwell
superfield strength. It enables one to divide the original system
of superfield equations into the pure kinematical and dynamical parts 
which are separately
invariant under the original hidden supersymmetry, and to construct
the correct $N=2$ superfield action yielding the dynamical part
as the equation of motion. In this way we reconstructed the
$N=2$ BI action with the hidden extra $N=2$ supersymmetry up to
the sixth order in $N=2$ Maxwell superfield strength.

Although this approach can in principle be applied to restore, step by step,
the $N=4 \rightarrow N=2$ BI action to any order, it would be desirable to
develop a more direct way for constructing such an action, similar to the
method which was used in \cite{BG} in the $N=2 \rightarrow N=1,\, d=4$ 
case (see also \cite{RT} and \cite{IK}).
It deals with the linear Maxwell superfield strength from the very
beginning, and it is based upon completing the latter to a linear off-shell
supermultiplet of the full supersymmetry by adding a few extra superfields
of the unbroken supersymmetry. After imposing a nonlinear 
covariant constraint
on the superfields of the linear supermultiplet one ends up with
a nonlinear realization of the full supersymmetry in terms of
the Maxwell superfield strength as the only independent Goldstone superfield.
In all the cases studied so far, both the BI superfield Lagrangian density
and the Goldstone-Maxwell superfield strength belong to the linear
supermultiplet just mentioned.

In this paper we propose a generalization of the method of \cite{{BG},{RT}}
to the $N=4 \rightarrow N=2$ BI case. Two essentially novel closely related
points of our construction, as compared to the previously elaborated cases,
are as follows: (i) We start from a proper extension of $N=4,\, d=4$ Poincar\'e
superalgebra by {\it a complex central charge}, in order to gain a geometric
place for the complex bosonic $N=2$ Maxwell superfield strength ${\cal W}$
as the Goldstone superfield \cite{BIK3}; (ii) The minimal linear $N=4$
supermultiplet into which one can embed ${\cal W}$ necessarily involves an
{\it infinite} tower of chiral $N=2$ superfields of growing dimension
interrelated by the central charge generators. As in the previous cases,
the chiral $N=2$ Lagrangian density of the $N=4 \rightarrow N=2$ BI theory
is one of these extra superfields, but in order to express it in terms of
${\cal W}, \bar{\cal W}$ one is led to impose an {\it infinite} set of the
covariant constraints which eliminate all the extra superfields as well. We
give these constraints in the explicit form and solve them by iterations,
in order to
restore the correct $N=4 \rightarrow N=2$ BI action up to the 10th order in
${\cal W}, \bar{\cal W}$. Surprisingly, up to the 8th order it reproduces
the $N=2$ action found earlier to this order in \cite{KT} from
the requirements of self-duality and invariance under a shift
symmetry of ${\cal W}, \bar{\cal W}$ (in our approach it comes out as the
symmetry generated by central charges). This is an indication 
that the requirements
of \cite{KT} are equivalent to the single demand of hidden $N=2$
supersymmetry. As a consequence, the full 
$N=4 \rightarrow N=2, \;d=4$ BI action is
expected to be self-dual.

\setcounter{equation}0
\section{Getting started}

The basic object we shall deal with is a complex scalar
$N=2$ off-shell superfield strength ${\cal W}$. It is chiral and
satisfies one additional Bianchi identity
\be
(a)\;\bar D_{\dot\alpha i} {\cal W} =0~,\;
D^i_{\alpha} \bar{\cal W} =0~,\quad (b)\;
D^{ik}{\cal W} = \bar D^{ik}\bar{\cal  W}~. \label{vector}
\ee
Here
\bea
&&D_{\alpha}^i = \frac{\partial}{\partial\theta^{\alpha}_i}+
  i{\bar\theta}^{{\dot\alpha}i}\partial_{\alpha\dot\alpha}\; ,\quad
{\bar D}_{{\dot\alpha}i}=-\frac{\partial}
{\partial{\bar\theta}^{\dot\alpha i}}
  -i\theta^{\alpha}_i\partial_{\alpha\dot\alpha}\;, \quad
 \left\{ D_{\alpha}^i, {\bar D}_{{\dot\alpha}j}\right\}=
 -2i\delta^i_j \partial_{\alpha\dot\alpha}~,\label{cd} \\
&& D^{ij} \equiv  D^{\alpha\, i}D_{\alpha}^j~, \quad
\bar D^{ij} \equiv  \bar D^{i}_{\dot\alpha}\bar D^{\dot\alpha \,j}~. \lb{defDD}
\eea
Due to the basic constraints \p{vector} the superfields $\cW,\cWb$
obey the following useful relations:
\be
D^4 \cW = -\frac{1}{2} \Box \cWb \;, \quad {\bar D}{}^4 \cWb 
= -\frac{1}{2}\Box \cW \;,
\ee
where
\bea
D^4\equiv \frac{1}{48} D^{\alpha i}D^j_\alpha
D^{\beta}_{ i}D_{\beta j} \; , \quad \bar D^4 =
\overline{(D^4)} \equiv\frac{1}{48}
\bar D_{\dot\alpha}^i \bar D^{\dot\alpha j}
   \bar D_{\dot\beta i}\bar D^{\dot\beta}_ j~ \; ,\quad
\Box \equiv
\partial_{\alpha\dot\alpha}\partial^{\alpha\dot\alpha} \;.
 \label{def1}
\eea
The irreducible field content of ${\cal W}$, $\bar{\cal W}$ 
defined by \p{vector}
is the off-shell $N=2,\, d=4$ abelian vector multiplet. 
It consists of a $SU(2)$-singlet complex
scalar field, a Maxwell field strength, a real $SU(2)$-triplet 
of the scalar bosonic auxiliary fields
and a $SU(2)$-doublet of Weyl spinors, i.e. a total of $(8+8)$ 
independent components.

In order to be able to treat ${\cal W}, \cWb$  as the
Goldstone superfields associated with the PBGS pattern
$N=4 \rightarrow N=2$ in $d=4$ we should define, first of all,
an appropriate modification of the standard $N=4,\, d=4$ Poincar\'e
superalgebra. It should involve a complex bosonic generator to which
one could put in correspondence  ${\cal W}, \cWb$ as the Goldstone
superfields. As was shown in \cite{BIK3}, the proper extension is
given by the following superalgebra:
\bea
&& \left\{ Q_{\alpha}^i, {\bar Q}_{\dot{\alpha}j}
        \right\}=2\delta^i_jP_{\alpha\dot{\alpha}} \;, \quad
\left\{ S_{\alpha}^i, {\bar S}_{\dot{\alpha}j}
        \right\}=2\delta^i_jP_{\alpha\dot{\alpha}}\;, \nn
&& \left\{ Q_{\alpha}^i,  S_{\beta}^j
   \right\}=2\varepsilon^{ij}\varepsilon_{\alpha\beta}Z \;, \quad
      \left\{ {\bar Q}_{\dot{\alpha}i}, {\bar S}_{\dot{\beta}j}
        \right\}=
 -2\varepsilon_{ij}\varepsilon_{\dot{\alpha}\dot{\beta}}{\bar Z} \;,
\;\; (i, j = 1,2)~, \label{n4sa}
\eea
with all other (anti)commutators vanishing. It was chosen in \cite{BIK3} as
the starting point for constructing a nonlinear realization of the
PBGS pattern $N=4 \rightarrow N=2$, with the generators
$Q_{\alpha}^i, {\bar Q}_{\dot{\alpha}j}$, $P_{\alpha\dot{\alpha}}$
corresponding to the unbroken $N=2$ supersymmetry and the remaining ones to
the spontaneously broken symmetries. Obviously, a linear realization of the
same version of PBGS should proceed from the
same superalgebra. The necessary presence of the complex central charge $Z$
in the anticommutators of the broken and unbroken $N=2$ spinor
generators is the crucial difference of the case under consideration from
the PBGS case $N=2 \rightarrow N=1$ in $d=4$ \cite{BG} and its $d=3$
counterpart \cite{IK}. In the latter two cases one proceeds from the
$N=2$ Poincar\'e supersymmetries in $d=4$ and $d=3$ with no central charge
generators; the elementary Goldstone superfields are the fermionic ones which
are eventually identified with the corresponding $N=1$ Maxwell 
superfield strengths.
Note that the superalgebra \p{n4sa} is a $d=4$ notation for the
$N=(2,0)$ (or $N=(0,2)$) Poincar\'e superalgebra in $D=6$. In what follows,
we shall not actually need to resort to the $D=6$ interpretation. 
We shall entirely deal with
the $N=2,\, d=4$  superfields, viewing \p{n4sa} as 
a central-charge extension
of the standard $N=4$ Poincar\'e supersymmetry in $d=4$.

\setcounter{equation}0
\section{Embedding N=2 vector multiplet into a linear N=4 multiplet}
In \cite{BIK3}, starting from a nonlinear realization of $N=4$ supersymmetry
defined by the superalgebra \p{n4sa}, we found, up to the 4th order in
fields, an equivalence transformation from the $N=2$ Goldstone superfield
associated with the generator $Z$ to the standard $N=2$ Maxwell superfield
strength ${\cal W}, \bar{\cal W}$ defined by the constraints \p{vector}.
We found that the nonlinear hidden $N=2$ supersymmetry and 
$Z, \bar Z$ symmetry
are realized on ${\cal W}, \bar{\cal W}$ by the following transformations:
\be
\delta {\cal W} = f\left(1 -{1\over 2}\bar D^4 {\cAb}_0 \right) +
{1\over 4} \bar f \Box \cA_0  + {1\over 4i}
\,\bar D^{i\dot\alpha}\bar f 
D^{\alpha}_i\,\partial_{\alpha\dot\alpha} \cA_0,
 \quad \delta \bar{\cal W} = (\delta {\cal W})^*~,
\label{N4a}
\ee
where the functions $f,\bar f$,
\be
f = c+ 2i\,\eta^{i\alpha}\theta_{i\alpha}~, \quad \bar f =
\bar c  - 2i\,\bar\eta^i_{\dot\alpha} \bar\theta^{\dot\alpha}_i~,
\label{fdef}
\ee
collect the parameters of broken supersymmetry
($\eta^{i\alpha},\bar\eta^i_{\dot\alpha}$)
and those of the central charge transformations ($c,\bar c$). 
The complex chiral function
${\cal A}_0$ was specified up to the fourth order
\footnote{For further convenience, here we use a
slightly different notation for this function as compared 
to ref. \cite{BIK3}.}
\bea
&&{\cal A}_0 = 
{\cal W}^2\left(1 + {1\over 2} \bar D^4 \bar{\cal W}^2 \right) +
O({\cal W}^6)~,  \label{init} \\
&& \bar D_{\dot\alpha i} {\cal A}_0 =0~. \label{chirA0}
\eea

Actually, the transformation law \p{N4a}, \p{fdef} is the most general
hidden supersymmetry transformation law of ${\cal W}, \bar{\cal W}$
compatible with the defining constraints \p{vector}, provided that the
$N=2$ superfield function ${\cal A}_0$ obeys the chirality 
condition \p{chirA0}.
By analogy with the $N=1$ construction of \cite{BG}, in order to promote
\p{N4a} to a {\it linear} (though still inhomogeneous)
realization of the considered $N=4$ supersymmetry, it is natural
to treat ${\cal A}_0$ as a new {\it independent} $N=2$ superfield
constrained only by the chirality condition \p{chirA0} and to try to
define the transformation law of ${\cal A}_0$ under the $\eta, \bar \eta,
c, \bar c$-transformations in such a way that the 
$N=2$ superfields ${\cal A}_0$,
${\cal W}, \bar{\cal W}$ form a closed set. Then, imposing a proper
covariant constraint on these superfields one could hope to recover
the structure \p{init} as a solution to this constraint. In view of the
covariance of this hypothetical constraint, the correct transformation law
for ${\cal A}_0$ to the appropriate order can be reproduced by
varying \p{init} according to the transformation law \p{N4a}. 
Since we know ${\cal A}_0$
up to the 4th order, we can uniquely restore its transformation law
up to the 3d order. We explicitly find
\be
\delta {\cA_0} = 2f\cW  +
{1\over 4} \bar f \Box \cA_1 + {1\over 4i}
\,\bar D^{i\dot\alpha}\bar f 
D^{\alpha}_i\,\partial_{\alpha\dot\alpha} \cA_1,
\label{probe}
\ee
where
\be
{\cal A}_1 = {2\over 3}\,{\cal W}^3 + O({\cal W}^5)~, 
\quad \bar D_{\dot\alpha i} {\cal A}_1 =0~.
\ee
We observe the appearance of a new composite chiral superfield ${\cal A}_1$,
and there is no way to avoid it in the transformation law \p{probe}. This is
the crucial difference from the $N=1$ case of ref. \cite{{BG},{RT}} where
a similar reasoning led to a closed supermultiplet with only one extra
$N=1$ superfield besides the $N=1$ Goldstone-Maxwell one (the resulting
linear multiplet of $N=2$ supersymmetry is a $N=1$ superfield form
of the $N=2$ vector multiplet with a modified transformation 
law \cite{part,ibmz}).

Thus, we are forced to incorporate a chiral superfield ${\cal A}_1$ as a new
independent $N=2$ superfield component of the linear $N=4$ supermultiplet we
are seeking. Inspecting the brackets of all these transformations
suggests that the only possibility to achieve their closure in
accord with the superalgebra \p{n4sa} is to introduce an {\it infinite}
sequence of chiral $N=2$ superfields and their antichiral conjugates
$\cA_n\;,~{\cAb}_n$, $n = 0, 1, \ldots$,
\be
\bar D_{\dot\alpha i} {\cA_n} =0~,\;
D^i_{\alpha} \bar{\cA_n} =0~, \label{defA}
\ee
with the following transformation laws:
\bea
&& \delta {\cA_0} = 2f\cW  +
{1\over 4}\bar f \Box \cA_1 + {1\over 4i}
\,\bar D^{i\dot\alpha}\bar f D^{\alpha}_i\,
\partial_{\alpha\dot\alpha} \cA_1, \label{N4b1} \\
&&
\delta {\cA_1} = 2f \cA_{0}  +
{1\over 4}\bar f \Box \cA_{2} + {1\over 4i}
\,\bar D^{i\dot\alpha}\bar f D^{\alpha}_i\,\partial_{\alpha\dot\alpha}
\cA_{2} \nn
&& ........... \nonumber \\
&& \delta {\cA_n} = 2f \cA_{n-1}  +
{1\over 4} \bar f \Box \cA_{n+1} + {1\over 4i}
\,\bar D^{i\dot\alpha}\bar f D^{\alpha}_i\,
\partial_{\alpha\dot\alpha} \cA_{n+1},\quad (n\geq 1)\label{N4b} \\
&& \delta \bar{\cA_n} = (\delta {\cA}_n)^*~ \nonumber
 \;.
\eea
It is a simple exercise to check that these transformations 
close off shell both among
themselves and with those of the manifest $N=2$ supersymmetry 
just according to
the $N=4$ superalgebra \p{n4sa}.

Realizing (formally) the central charge generators
as derivatives in some extra complex ``central-charge coordinate'' $z$
\be
Z = {i\over 2} \frac{\partial}{\partial z}~, \quad \bar Z = {i\over 2}
\frac{\partial}{\partial \bar z}~, \label{defz}
\ee
and assuming all the
involved $N=2$ superfields to be defined on a $z, \bar z$ extension of the
standard $N=2$ superspace, it is instructive to rewrite the transformation
laws under the $c, \bar c$ transformations as follows:
\bea
\frac{\partial {\cal W}}{\partial z} = 
\left( 1-\frac{1}{2}{\bar D}^4 \bar{\cal A}_0
\right)\;, \quad  \frac{\partial {\cal W}}{\partial\bar z}
=\frac{1}{4}\Box {\cal A}_0\;,
\label{cc1} \\
\frac{\partial {\cal A}_0}{\partial z} = 2{\cal W}\;, \quad \frac{\partial
{\cal A}_0}{\partial\bar z}= \frac{1}{4}\Box {\cal A}_1 \;, \label{cc2} \\
\frac{\partial
{\cal A}_n}{\partial z} = 2{\cal A}_{n-1}\;, 
\quad \frac{\partial {\cal A}_n}{\partial\bar z}=
\frac{1}{4}\Box {\cal A}_{n+1} \;. \label{cc}
\eea
These relations imply, in particular,
\bea
\left(\frac{\partial^2}{\partial z\partial {\bar
z}} -\frac{1}{2} \Box \right) {\cal W}=0\; , \quad
\left(\frac{\partial^2}{\partial z\partial {\bar z}} -\frac{1}{2} \Box
\right) {\cal A}_n=0~. \label{mass}
\eea
If we regard $z, \bar z$ as the actual coordinates,
which extend the $d=4$ Minkowski
space to the $D=6$ one, the relations \p{mass} mean that the constructed
linear supermultiplet is on shell from the $D=6$ perspective. On the other
hand, from the $d=4$ point of view this multiplet is off-shell, and the
relations \p{cc1} - \p{cc}, \p{mass} simply give 
a specific realization of the central
charge generators $Z, \bar Z$ on its $N=2$ superfield components. In this
sense this multiplet is similar to the previously known special $N=2,\, d=4$
and $N=4,\, d=4$ supermultiplets, which are obtained from the on-shell
multiplets in higher dimensions via non-trivial dimension reductions and
inherit the higher-dimensional translation generators as non-trivially
realized central charges in $d=4$ \cite{FS,SSW} 
(a renowned example of this sort
is the $(8+8)$ Fayet-Sohnius hypermultiplet \cite{FS}). Since the
superalgebra \p{n4sa} is just a $d=4$ form of the $N=(2,0)$ (or $N=(0,2)$)
$D=6$ Poincar\'e superalgebra, it is natural to think that
the above supermultiplet has a $D=6$
origin and to try to reveal it.\footnote{By analogy with the previously
known examples \cite{BG,IK}, one could expect, at first glance, that this
multiplet is a $d=4$ form of the vector $N=2,\, D=6$ multiplet which is known
to exist only on shell (in $D=6$) \cite{D6}. However, this cannot be true
because such a multiplet can be defined only for the $N=(1,1)$ supersymmetry
in $D=6$ \cite{D6} while we are facing $N=(2,0)$ or $N=(0,2)$ supersymmetry
in our case. Note that the PBGS option $N=4 \rightarrow N=2,\, d=4$,
with the $N=4,\, d=4$ supersymmetry being isomorphic 
just to the $N=(1,1),\, D=6$ one,
was discussed in \cite{BIK1}. It requires the $N=2$ hypermultiplet, as the
Goldstone multiplet.} We hope to come back to this interesting 
problem in the
future. For the time being we prefer to treat the above 
infinite-dimensional
representation in the pure $d=4$ framework 
as a linear realization of the partial
spontaneous breaking of the central-charge extended $N=4,\, d=4$
supersymmetry \p{n4sa} to the standard $N=2$ supersymmetry. 
The Goldstone
character of this realization is manifested in the transformation
law \p{N4a} which contains pure shifts by the parameters of the
spontaneously broken symmetries. Therefore, the appropriate components of
the superfield strength ${\cal W}$ are the Goldstone fields, and this
superfield itself can be interpreted as the Goldstone $N=2$ superfield of
the linear realization of the considered 
PBGS pattern $N=4 \rightarrow N=2$.

\setcounter{equation}{0}
\section{Superfield action of the $N=4 \rightarrow N=2$ BI theory}
As was already mentioned, in the approach proceeding 
{}from a linear realization of PBGS,
the Goldstone superfield Lagrange density is, as a rule,
a component of the same linear supermultiplet to which 
the relevant Goldstone
superfield belongs. This is also true for the case under consideration.
A good candidate for the chiral $N=2$ Lagrangian density 
is the superfield $\cA_0$.
Indeed, the ``action''
\be
S=\int d^4x d^4\theta \cA_0 + \int d^4x d^4\bar\theta \cAb_0 \label{action1}
\ee
is invariant with respect to the transformation \p{N4b1} up to surface
terms, because, taking into account the basic constraints \p{vector} and the
precise form of these transformations, the integrand is shifted by
$x$-derivatives. With the interpretation of the central charge
transformations as shifts with respect to the coordinates $z, \bar z$, the
action \p{action1} does not depend on these coordinates in virtue of eqs.
\p{cc2}, though the Lagrangian density can bear such a dependence.

It remains for us to define covariant constraints 
which would express $\cA_0$,  $\cAb_0$ in terms
of ${\cal W}$, $\bar{\cal W}$, with preserving the linear representation
structure \p{N4a}, \p{N4b1}, \p{N4b}. Because an infinite number
of $N=2$ superfields ${\cal A}_n$ is present in our case,
there should exist an infinite set of constraints trading 
all these superfields
for the basic Goldstone ones ${\cal W}$, $\bar{\cal W}$.

As a first step in finding these constraints let us note that
the following expression:
\be\label{constr1}
\phi_0 = \cA_0 \left( 1-\frac{1}{2}{\bar D}{}^4\cAb_0\right) -
\cW^2- \sum_{k=1}\frac{ (-1)^k}{2\cdot 8^k}
   \cA_k\Box^k {\bar D}{}^4 \cAb_k
\ee
is invariant, with respect to the $f$ part of
the transformations \p{N4a}, \p{N4b1} - \p{N4b}.
This leads us to choose
\be\label{constr1a}
\phi_0=0
\ee
as our first constraint. For consistency with $N=4$ supersymmetry,
the constraint \p{constr1a} should be invariant with respect
to the full transformations \p{N4a}, \p{N4b1}, \p{N4b}, with the
$\bar f$ part taken into account as well. We shall firstly specialize
to the $\bar c$ part of the $\bar f$ transformations. 
The requirement of the $\bar c$ covariance produces 
the new constraint
\be\label{constr2}
\phi_1=\Box \cA_1 +2\left( \cA_0\Box \cW - \cW\Box\cA_0 \right)
-\sum_{k=0}\frac{ (-1)^k}{2\cdot8^k}
\left( \Box \cA_{k+1}\Box^k{\bar D}{}^4 \cAb_k-
  \cA_{k+1} \Box^{k+1} {\bar D}{}^4 \cAb_k    \right) =0\;.
\ee
It is invariant under the $f$ transformations, but requiring it
to be invariant also under the $\bar c$ part gives rise 
to the new constraint
\bea\label{constr3}
\phi_2 &=&\Box^2 \cA_2 +2\left( \cA_0\Box^2 \cA_0 
- \Box\cA_0\Box\cA_0+2\Box\cA_1\Box\cW -
   \cA_1\Box^2\cW -\cW\Box^2\cA_1 \right)  \nonumber \\
&&-\sum_{k=0}\frac{ (-1)^k}{2\cdot8^k}
\left( \Box^2 \cA_{k+2}\Box^k{\bar D}{}^4 \cAb_k-
     2\Box \cA_{k+2}\Box^{k+1}{\bar D}{}^4 \cAb_k +
    \cA_{k+2} \Box^{k+2} {\bar D}{}^4 \cAb_k    \right) =0\,.
\eea
Applying the same procedure to \p{constr3}, we find the next constraint
\bea\label{constr4}
\phi_3 &=&\Box^3 \cA_3 +2\left( 3\Box^2\cA_2\Box\cW 
+ 3\Box\cA_1\Box^2\cA_0+\cA_0\Box^3\cA_1-\cA_1\Box^3\cA_0
 +\cA_2\Box^3\cW \right. \nn
&& -\left.\cW\Box^3\cA_2 -3\Box^2\cA_1\Box\cA_0
-3\Box\cA_2\Box^2\cW\right)+ \ldots =0 ,
\eea
and so on. The full infinite set of constraints is by construction
invariant under the $f$ and $\bar c$ transformations. Indeed,
using the relations \p{cc1}-\p{cc} one may
explicitly check that
\be
\frac{\partial \phi_n}{\partial z} = 0~, \quad
\frac{\partial \phi_n}{\partial \bar z} = {1\over 4}\phi_{n+1}~,
\ee
so the full set of constraints is indeed closed.

The variation of the basic constraints with respect to
the $\bar f$ transformations has the following general form:
\be
\delta \phi_n = \bar\eta{}^{i\dot\alpha}{\bar\theta}_{i\dot\alpha} 
{\cal B}_n + \bar\eta{}^{i\dot\alpha} ({\cal F}_n)_{i\dot\alpha} \;.
\ee
Demanding this variation to vanish gives rise to the two sets of constraints
\be\label{cccc}
(\mbox{a})\;\; {\cal B}_n=0 \;, \quad (\mbox{b})\;\; 
({\cal F}_n)_{i\dot\alpha} =0 \;.
\ee
The constraints (\ref{cccc}a) are easily recognized as those 
obtained above from
the $\bar c$ covariance reasoning. One can show by explicit 
calculations that
\be
{\bar D}{}^{i\dot\alpha} ({\cal F}_n)_{j\dot\beta} \sim \delta^i_j
    \delta^{\dot\alpha}_{\dot\beta} {\cal B}_n \;.
\ee
Thus the fermionic constraints (\ref{cccc}b) seem to be more 
fundamental. For example, for the
constraint $\phi_1$ \p{constr2} the basic fermionic constraint reads
\bea\label{constr2f}
({\cal F}_1)_{i\dot\alpha} &=& 
D^\alpha_i\partial_{\alpha\dot\alpha}\cA_1 +
2\left( \cA_0 D^\alpha_i\partial_{\alpha\dot\alpha} \cW 
- \cW D^\alpha_i\partial_{\alpha\dot\alpha}\cA_0 \right) \nn
&& -\sum_{k=0}\frac{ (-1)^k}{2\cdot 8^k}
\left( D^\alpha_i\partial_{\alpha\dot\alpha} 
\cA_{k+1}\Box^k{\bar D}{}^4 \cAb_k-
  \cA_{k+1} \Box^{k}D^\alpha_i\partial_{\alpha\dot\alpha} 
{\bar D}{}^4 \cAb_k    \right) =0 \;.
\eea
In order to prove that the basic fermionic constraints 
(\ref{cccc}b) are actually equivalent to the
bosonic ones (\ref{cccc}a), one has to know the 
general solution to {\it all} constraints.
For the time being we have explicitly checked this 
important property only for the iteration solution
given below. Taking for granted that this is true in general, 
we can limit our
attention to the type $(\mbox{a})$ constraints only. The constraints
\p{constr1},
\p{constr2} are just of this type.

At present we have no idea, how to explicitly solve
the above infinite set of constraints and find a closed
expression for the Lagrangian densities ${\cal A}_0$, $\bar{\cal A}_0$
similar to the one known in the $N=2 \rightarrow N=1$ case \cite{BG}.
What we are actually able to do, so far, is to
restore a general solution by iterations. E.g., in order to restore
the action up to the 10th order, we have to know
the following orders in $\cA_k$:
\bea
&& \cA_0= \cW^2 + \cA_0^{(4)}+\cA_0^{(6)}+\cA_0^{(8)}+\ldots \; ,\nn
&& \cA_1= \cA_1^{(3)}+\cA_1^{(5)}+\cA_1^{(7)}+\ldots \; , \nn
&& \cA_2= \cA_2^{(4)}+\cA_2^{(6)}+\ldots \;, \quad \cA_3
= \cA_3^{(5)}+\ldots \; .
\eea
These terms were found to have the following explicit structure:
\bea\label{An}
\cA_0^{(4)}&=& \frac{1}{2}\cW^2 {\bar D}{}^4\cWb^2\;, \quad
  \cA_0^{(6)}=\frac{1}{4}{\bar D}{}^4 
\left[ \cW^2\cWb^2\left( D^4\cW^2+{\bar D}{}^4\cWb^2\right)-
   \frac{1}{9}\cW^3\Box\cWb^3\right]\;, \nn
\cA_0^{(8)}& =&\frac{1}{8}{\bar D}{}^4\left[ 4\cW^2\cAb_0^{(6)} 
+ 4\cWb^2\cA_0^{(6)} +
  \cW^2\cWb^2D^4\cW^2{\bar D}{}^4\cWb^2
-\frac{2}{9}\cW^3\Box\left(\cWb^3D^4\cW^2\right) \right. \nn
&& \left. -\frac{2}{9}\cW^3{\bar D}{}^4\cWb^2\Box\cWb^3
+\frac{1}{144}\cW^4\Box^2\cWb^4 \right] \;, \nn
\cA_1^{(3)}& = & \frac{2}{3}\cW^3\;, \quad \cA_1^{(5)}
=\frac{2}{3}\cW^3{\bar D}{}^4\cWb^2 \;, \nn
\cA_1^{(7)}&=& {\bar D}{}^4
\left[ \frac{1}{2}\cW^3\cWb^2{\bar D}{}^4\cWb^2+
   \frac{1}{3}\cW^3\cWb^2D^4\cW^2
-\frac{1}{24}\cW^4\Box\cWb^3 \right] \;, \nn
\cA_2^{(4)}& = & \frac{1}{3}\cW^4\;,\quad \cA_2^{(6)} 
=  \frac{1}{2}\cW^4{\bar D}{}^4\cWb^2\;,
 \quad \cA_3^{(5)}=\frac{2}{15}\cW^5 \;.
\eea

Note that, despite the presence of growing powers 
of the operator $\Box$ in our constraints,
in each case the maximal power of $\Box$ can be finally taken off 
{}from all the terms in the given
constraint, leaving us with this maximal power of $\Box$ acting on
an expression which starts from the appropriate ${\cal A}_n$. 
Equating these final expressions
to zero allows us to algebraically express all ${\cal A}_n$ 
in terms of ${\cal W}, \bar{\cal W}$
and derivatives of the latter. For example, 
for $\cA_3^{(5)}$ we finally get the following equation:
\be
\Box^3 \cA_3^{(5)} =\frac{2}{15} \Box^3 \cW^5 \; \Rightarrow \;
\cA_3^{(5)} =\frac{2}{15} \cW^5~.
\ee

This procedure of taking off the degrees of $\Box $ with discarding
possible ``zero modes'' can be justified as follows: we are interested
in an off-shell solution that preserves the manifest standard
$N=2$ supersymmetry including the Poincar\'e covariance.
This rules out possible on-shell
zero modes as well as the presence of explicit $\theta$'s or $x$'s
in the expressions which remain after taking off
the appropriate powers of $\Box$.
It can be checked to any desirable order that these ``reduced''
constraints yield correct local expressions 
for the composite superfields $\cA_n$,
which prove to transform just in accordance with the original 
transformation rules
\p{N4a}, \p{N4b1}, \p{N4b}. We have explicitly verified 
this for our iteration
solution \p{An}. We do not know, for the time being, how to demonstrate
the possibility to take off the powers of
$\Box $ from the original constraints in general, without explicitly
solving them. In Appendix we deduce a set of purely algebraic
constraints which immediately give the above iteration solution and so are
candidates for the general form of the ``reduced'' constraints.

The explicit expression for the action, up to the 8th order in
${\cal W}, \bar{\cal W}$, reads
\bea
&&S^{(8)} = \left(\int d^4 x d^4\theta \cW^2+\mbox{c.c.}\right)+
\int d^4 x d^4\theta d^4\bar\theta  \left\{ \cW^2\cWb^2\left[
 1+\frac{1}{2}\left( D^4\cW^2 +{\bar D}{}^4\cWb^2\right)\right] \right. \nn
&& -\frac{1}{18}\cW^3\Box\cWb^3
+\frac{1}{4} \cW^2\cWb^2\left[ \left( D^4\cW^2
+{\bar D}{}^4\cWb^2\right)^2+ D^4\cW^2{\bar D}{}^4\cWb^2
  \right] \nn
&& \left.-\frac{1}{12}D^4\cW^2\cWb^3\Box\cW^3
-\frac{1}{12}{\bar D}{}^4\cWb^2\cW^3\Box\cWb^3
+\frac{1}{576}\cW^4 \Box^2 \cWb^4 \right\} \;. \label{8}
\eea
This action, up to a slight difference in the notation, coincides with the
action found by Kuzenko and Theisen \cite{KT} from the requirements of
self-duality and invariance under nonlinear shifts of ${\cal W}, \bar{\cal
W}$ (the $c, \bar c$ transformations in our notation). Let us point out that
the structure of nonlinearities in the $c, \bar c$ transformations of
${\cal W}, \bar{\cal W}$ in our approach is uniquely fixed 
by the original $N=4$
supersymmetry transformations and the constraints imposed. In \cite{KT} it
was guessed order by order from the requirement that
the action be invariant.

The next, 10th order part of the $N=4$ invariant $N=2$ BI action 
can be easily restored
{}from eqs. \p{An}. Its explicit form looks not too enlightening,
so here we present only the relevant part of the chiral Lagrangian 
density  in the condensed notation
\bea
{\cal A}_0^{(10)}&=&\frac{1}{2}{\bar D}{}^4 \left[ \cW^2\cAb_0^{(8)} 
+\cWb{}^2\cA_0^{(8)}+
\cA_0^{(4)}\cAb_0^{(6)} +\cAb_0^{(4)}\cA_0^{(6)}
- \frac{1}{8} \cA_1^{(3)}\Box\cAb_1^{(7)}
- \frac{1}{8} \cA_1^{(7)}\Box\cAb_1^{(3)}\right. \nn
&& \left.
-\frac{1}{8}\cA_1^{(5)}\Box\cAb_1^{(5)}
+\frac{1}{64}\cA_2^{(4)}\Box^2\cAb_2^{(6)} +
 \frac{1}{64}\cA_2^{(6)}\Box^2\cAb_2^{(4)}
  -\frac{1}{512}\cA_3^{(5)}\Box^3\cAb_3^{(5)}\right]~.
\eea
It would be interesting to compare it with the 10th order 
of the Kuzenko-Theisen
action (which, unfortunately,
was not explicitly given in \cite{KT}). Anyway, the coincidence
of the action of \cite{KT} with the $N=4 \rightarrow N=2$ BI action,
up to the 8th order,
can be regarded as a strong indication that these actions coincide at any
order and, hence, that the $N=4 \rightarrow N=2$ BI action is self-dual 
like its
$N=2 \rightarrow N=1$ prototype \cite{BG,RT}.

Finally, let us point out that after doing the $\theta $ integral,
the pure Maxwell field strength part of the
bosonic sector of the above action (and of the hypothetical
complete action) comes entirely from the expansion
of the standard Born-Infeld bosonic action. Just in this sense the
above action is a particular $N=2$ extension of the bosonic BI action.
The difference from the action of ref. \cite{Ke} is just in 
higher-derivative
terms with the $\Box$ operators.
These correction terms are crucial for the invariance
under the hidden $N=2$ supersymmetry, and they
drastically change, as compared to ref. \cite{Ke}, the structure
of the bosonic action, both in the pure scalar fields sector and the
mixed sector involving couplings
between the Maxwell field strength and the
scalar fields. By a reasoning
of \cite{BIK3}, the additional terms are just those needed
for the existence of an equivalence field
redefinition bringing the scalar fields action into the
standard static-gauge Nambu-Goto form.

The analysis of the auxiliary field sector shows that the equation
for the auxiliary field $P^{(ik)}(x)$ has the following generic structure:
$$
P^{(ik)}M_{(ik)}^{(nl)} = 0~,
$$
where $M$ is a non-singular matrix, $M = I + \ldots $, 
and ``dots'' stand for
terms involving fields and their derivatives. 
Hence, $P^{(ik)} = 0$ on shell, i.e. the
auxiliary field is non-propagating, as in the standard 
$N=2$ Maxwell theory.

\section{Conclusion}
In this paper we proposed a systematic way of constructing a $N=2$
superfield BI action with a
hidden second $N=2$ supersymmetry.  It is based on
extending the $N=2$ vector multiplet to an infinite-dimensional linear
off-shell multiplet of the central-charge modified $N=4$ supersymmetry 
and imposing
an infinite set of covariant constraints which give rise to a nonlinear
realization of the $N=4$ supersymmetry in terms of the $N=2$ Maxwell
(Goldstone-Maxwell) superfield strengths ${\cal W}, \bar{\cal W}$. 
Solving these
constraints by iterations, we have restored the $N=4$ 
supersymmetric BI action
to 10th order in ${\cal W}, \bar{\cal W}$. In order
to construct the full action, 
we need
to know the general solution of the constraints. For this purpose it seems
necessary to work out another, technically more feasible way
of tackling the infinite set of these
constraints, perhaps in a $z, \bar z$ extended $N=4$ superspace,
rather than
in the $N=2$ one. We hope to report soon on a progress
in this direction. Another project for a future study is to apply
our approach to construct a genuine non-abelian version
of the $N=4 \rightarrow N=2$ BI action as a proper modification
of the action proposed in \cite{Ke1}.

In the course of writing this paper we learned that a construction
conceptually closed to ours was independently worked out by A. Galperin
\cite{sasha}.

\vspace{0.3cm}
\noindent{\bf Acknowledgements.}\hskip 1em
This work was supported in part by the Fondo Affari Internazionali 
Convenzione Particellare
INFN-JINR, grants RFBR-CNRS 98-02-22034, RFBR-DFG-99-02-04022,
RFBR 99-02-18417 and NATO Grant PST.CLG 974874.

\setcounter{equation}{0}
\def\theequation{A.\arabic{equation}}
\section*{Appendix}
Let us consider the following constraint
\be
\varphi_1 = \cA_1\left( 1- \frac{1}{2}{\bar D}{}^4\cAb_0\right) 
-\frac{2}{3}\cW\cA_0-
 \sum_{k=1}\frac{ (-1)^k}{8^k}\left( \frac{k}{3}+\frac{1}{2}\right)
   \cA_{k+1}\Box^k {\bar D}{}^4 \cAb_k  = 0\; .\label{constra}
\ee
Using \p{cc1},\p{cc2},\p{cc} and the following useful relation:
\be
\frac{\partial}{\partial z} \left( \sum_{k=1}\frac{ (-1)^k}{8^k} a_k
   \cA_{k+m}\Box^{k+p} {\bar D}{}^4 \cAb_{k+n} \right) 
= \frac{1}{4}\sum_{k=0}\frac{ (-1)^k}{8^k}\left( a_k-a_{k+1}\right)
   \cA_{k+m}\Box^{k+p+1} {\bar D}{}^4 \cAb_{k+n+1} \label{useful}
\ee
($a_0 \equiv 0$), one can easily check that
\be
\frac{\partial}{\partial z} \varphi_1 = \frac{4}{3} \phi_0 \; .
\ee
In other words, \p{constra} is the result of ``integrating'' the basic
constraint \p{constr1}, with respect to $z$. 
The same ``integration'' procedure can be continued
further to get the successive set of the algebraic constraints
\be\label{constr2a}
\varphi_2 = \cA_2\left( 1- \frac{1}{2}{\bar D}{}^4\cAb_0\right) 
-\frac{1}{2}\cW\cA_1-
 \sum_{k=1}\frac{ (-1)^k}{8^k}\left( \frac{k^2}{8}
+\frac{k}{2}+\frac{1}{2}\right)
   \cA_{k+2}\Box^k {\bar D}{}^4 \cAb_k = 0\; ,
\ee
$$\frac{\partial}{\partial z} \varphi_2 = \frac{3}{2} \varphi_1 \;, $$
\be\label{constr3a}
\varphi_3 = \cA_3\left( 1- \frac{1}{2}{\bar D}{}^4\cAb_0\right) 
-\frac{2}{5}\cW\cA_2-
 \sum_{k=1}\frac{ (-1)^k}{8^k}\left(\frac{k^3}{30}
+ \frac{k^2}{4}+\frac{37k}{60}+\frac{1}{2}\right)
   \cA_{k+3}\Box^k {\bar D}{}^4 \cAb_k = 0\; ,
\ee
$$\frac{\partial}{\partial z} \varphi_3 = \frac{8}{5} \varphi_2 \;, $$
and so on.

We have checked that the iteration solution of 
the constraints \p{constra}, \p{constr2a}, \p{constr3a},
up to the 8th order, exactly coincides with \p{An}, but we still 
have no general proof that this set of constraints
is indeed fundamental. It is, by construction, covariant under the
$c$- transformations and $\eta$-supersymmetry, but its covariance under the
$\bar f$ transformations remains to be proved. Note the interesting
relations betweeen the constraints \p{constr2} - \p{constr4} 
and \p{constra}, \p{constr2a}, \p{constr3a}
\be
\phi_1=\frac{\partial}{\partial {\bar z}} \phi_0 =
\frac{4}{3} \frac{\partial^2}{\partial z \partial {\bar z}}
\varphi_1~, \;\;
\phi_2 
= {32\over 9}\left(\frac{\partial^2}{\partial z \partial \bar z}\right)^2 
\varphi_2~, \;\;
\phi_3 = {80\over 9}\left(\frac{\partial^2}{\partial z \partial \bar
z}\right)^3\varphi_3~, \ldots~.
\ee


\begin{thebibliography}{99}
\bibitem{CF} S. Cecotti, S. Ferrara, Phys. Lett. {\bf B 187} (1987) 335.
\bibitem{Ts} A.A. Tseytlin, {\it Born-Infeld Action,
Supersymmetry and String Theory},
IMPERIAL-TP-98-99-67, Aug 1999; {\tt hep-th/9908105}.
\bibitem{BG} J. Bagger, A. Galperin, Phys. Rev. {\bf D 55} (1997) 1091.
\bibitem{RT} M. Ro\v{c}ek, A. Tseytlin, Phys. Rev. {\bf D 59} (1999) 106001.
\bibitem{IK} E. Ivanov, S. Krivonos, Phys. Lett. {\bf B 453} (1999) 237.
\bibitem{Ke} S. Ketov, Mod. Phys. Lett. {\bf A 14} (1999) 501;
Class. Quant. Grav. {\bf 17} (2000) L91.
\bibitem{BIK1} S. Bellucci, E. Ivanov, S. Krivonos,
 In: Proceedings of 32nd International Symposium Ahrenshoop,
 September 1 - 5, 1998, Buckow, Germany, Fortsch. Phys.
 {\bf 48} (2000) 19.
\bibitem{BIK2} S. Bellucci, E. Ivanov, S. Krivonos, 
Phys. Lett. {\bf B 460} (1999) 348.
\bibitem{KT} S.M. Kuzenko, S. Theisen,
{\it Nonlinear Selfduality and Supersymmetry},
LMU-TPW-00-19, Jul 2000; {\tt hep-th/0007231}; JHEP {\bf 0003} (2000) 034.
\bibitem{BIK0} S. Bellucci, E. Ivanov, S. Krivonos, Phys. Lett. {\bf B
482} (2000) 233.
\bibitem{BIK3} S. Bellucci, E. Ivanov, S. Krivonos, 
{\it N=2 and N=4 supersymmetric Born-Infeld theories
{}from nonlinear realizations}, LNF-00/040(P), JINR-E2-2000-311,
{\tt hep-th/0012236}, Phys. Lett. B (2001), to appear.
\bibitem{part} I. Antoniadis, H. Partouche, T.R. Taylor, 
Phys. Lett. {\bf B 372} (1996) 83.
\bibitem{ibmz} E.A. Ivanov, B.M. Zupnik, Yadern. Fiz. {\bf 62} (1999) 1110
({\tt hep-th/9710236}).
\bibitem{FS} M.F. Sohnius, Nucl. Phys. {\bf B 138} (1978) 109.
\bibitem{SSW} M.F. Sohnius, K.S. Stelle, P.C. West, 
Phys. Lett. {\bf 92 B} (1980) 123;
Nucl. Phys. {\bf B173} (1980) 127.
\bibitem{D6} P.S. Howe, G. Sierra, P.K. Townsend, Nucl. Phys. {\bf B 221}
(1983) 331.
\bibitem{Ke1} S. Ketov, Phys. Lett. {\bf B 491} (2000) 207.
\bibitem{sasha} A. Galperin, work in preparation.
\end{thebibliography}
\end{document}